# Dynamic Behaviors of Mix-game Model and Its Applications

Chengling Gou

Physics Department, Beijing University of Aeronautics and Astronautics

37 Xueyuan Road, Heidian District, Beijing, 100083, China

Physics Department, University of Oxford

Clarendon Laboratory, Parks Road, Oxford, OX1 3PU, UK

c.gou1@physics.ox.ac.uk, gouchengling@hotmail.com

**Abstract.** This paper proposes a modification to Minority Game (MG) by adding some agents who play majority game into MG. So it is referred to as mix-game. The highlight of this model is that the two groups of agents in mix-game have different bounded abilities to deal with history information and to count their own performance. Through simulations, this paper finds out that the local volatilities change a lot by adding some agents who play majority game into MG, and the change of local volatilities largely depends on different combinations of history memories of the two groups. Furthermore this paper analyses the underlying mechanisms for this finding. It also gives an example of applications of mix-game.

**Keyword:** minority game, majority game, mix-game,

## 1. Introduction

Minority game (MG) is an important agent-based financial market model proposed by D. Challet, and Y. C. Zhang [1]. Neil Johnson [2, 3] and coworkers extended MG by allowing a variable number of active traders at each timestep--- they called their modified game as the Grand Canonical Minority Game (GCMG). The GCMG, and to a lesser extent the basic MG itself, can reproduce the stylized facts of financial markets, such as volatility clustering and fat-tail distributions.

However, there are some weaknesses in MG and GCMG. First, the diversity of agents is limited, since agents all have the same memory and time-horizon. Second, in real markets, some agents are trend chasers, i.e. "noise traders" [4~11], who effectively play a majority game; while others are "fundamentalist", who effectively play a minority game. De Martino, I. Giardina1, M. Marsili and A.

Tedeschi have done some research on mixed Minority/Majority Games analytically and simulationally [19~23]. In their models, time-horizon is infinite and all agents have the same ability to deal with information and can switch between majority and minority. They study the stationary macroscopic properties and the dynamical features of the systems under different information structures, i.e. agents receive the information which is either exogenous ('random') or endogenous ('real'). They find that (a) a significant loss of informational efficiency with respect to a pure minority game (in particular, an efficient, unpredictable phase exists only if the fraction of agents in majority group < 1/2), and (b) a catastrophic increase of global fluctuations if the fraction of agents in majority group > 1/2 under the condition that agents receive random information [20]. They also find that a small amount of herding tendency can alter the collective behavior dramatically if agents receive endogenous information (real history) [22]. Zhong et al study a similar model in which they introduce contrarians into MG who deliberately prefer to hold an opinion that is contrary to the prevailing idea of the commons or normal agents. Their results of numerical simulations reveal that the average success rate among the agents depends non-monotonically on the fraction of contrarians [29].

In order to create an agent-based model which more closely mimics a real financial market, this paper modifies the MG model by dividing agents into two groups: each group has different memory and time-horizon; one group play the minority game and the other play the majority game. For this reason, I will refer to this system as a 'mix-game' model. The difference between mix-game and the mixed Minority/Majority Games studied by Marsili and Martino et al is that the two groups of agents in mix-game have different bounded abilities to deal with history information and to count their own performance. This feature of mix-game represents that agents have bounded rationality [5].

This paper looks at the dynamic features of mix-game under different combinations of history memories of the two groups and agents receiving real history information. The study focuses on the interplay between the strategy spaces of these two groups. Reference [32] reported that almost all agents play with memories of 6 or less in a typical minority game with evolution. This means memories of 6 or less is more important and worthwhile to look at. Therefore, I focus on studying the effect on MG by adding some agents playing majority game with history memories of 6 or less. Section 2 will introduces mix-game model and the simulation condition. Section 3 reports the simulation results and discussion. Section 4 gives examples of application of mix-game model. Section 5 reaches the conclusion of this paper.

## 2. The mix-game model and simulation condition

In mix-game, there are two groups of agents; group1 plays the majority game, and the group2 plays the minority game. N is the total number of the agents and N1 is number of agents of group1. The system resource is R=N*L, where L<1 is the fraction of the system resource. All agents compete in the system for the limited resource R. T1 and T2 are the time horizon lengths of the two groups of agents, and m1 and m2 denote the memory lengths of the two groups of agents, respectively.

Only the global information available to the agents is a common bit-string "memory" of the m1 or m2 most recent competition outcomes. A strategy consists of a response, i.e., 0 or 1, to each possible bit string; hence there are $2^{2^{m1}}$ or $2^{2^{m21}}$ possible strategies for group 1 or group 2, respectively, which form full strategy spaces (FSS). At the beginning of the game, each agent is assigned s strategies and keeps them unchangeable during the game. After each turn, agents assign one (virtual) point to a strategy which would have predicted the correct outcome. For agents in group 1, they will reward their strategies one point if they are in the majority; for agents in group 2, they will reward their strategies one point if they are in the minority. Agents collect the virtual points for their strategies over the time horizon T1 or T2, and agents use their strategy which has the highest virtual point in each turn. If there are two strategies which have the highest virtual point, agents use coin toss to decide which strategy to be used. Excess demand $D[(t)^-]$ is equal to the number of ones (buy) which agents choose minus the number of zeros (sell) which agents choose.

$$D[(t)^-] = n_{bur-orders}[t-1] - n_{sell-orders}[t-1] \quad (1)$$

According to a widely accepted assumption that excess demand exerts a force on the price of the asset and the change of price is proportion to the excess demand in a financial market [12, 13], the time series of price of the asset $P[(t)]$ can be calculated based on the time series of excess demand.

$$P[(t)] = D[t^-]/\lambda + P[t-1] \quad (2)$$

For simplicity, one usually chooses $\lambda = 1$. Volatility of the time series of prices is represented by the variance of the increases of prices. The local volatility Vol. is calculated at every time-step by calculating the variance of the increases of prices within a small time-step window $d$. If $\lambda = 1$, the increase of prices is just the excess demand. Therefore, we get formula (3).

$$Vol[t] = \frac{1}{d}\sum_{t-d}^{t} D[t^-]^2 - (\frac{1}{d}\sum_{t-d}^{t} D[t^-])^2 \quad (3)$$

In simulations, the distribution of initial strategies of agents is randomly uniform in FSS and remains unchanged during the game. Each agent has two strategies. The simulation running time-steps are 3000 and L is 0.5. The window length of local volatility is $d=5$.

## 3. Simulation results and discussions

**3.1 The effect of mix-game on local volatilities**

To see the effect of mix-game on local volatilities, simulations are done with different parameter configurations of memory lengths, number of total agents and fraction of agents in group1. In order to see the interplay between agent strategy spaces of the two groups, I choose three categories of configurations of memory lengths: (a) m1<m2=6; (b) m1=m1<=6; (c) m1=6>m2. Number of agents N varies from 101 to 401. Time horizons (T1 and T2) are relatively stable. I will look at the influence of time horizons (T1 and T2) on volatilities of mix-game in the future.

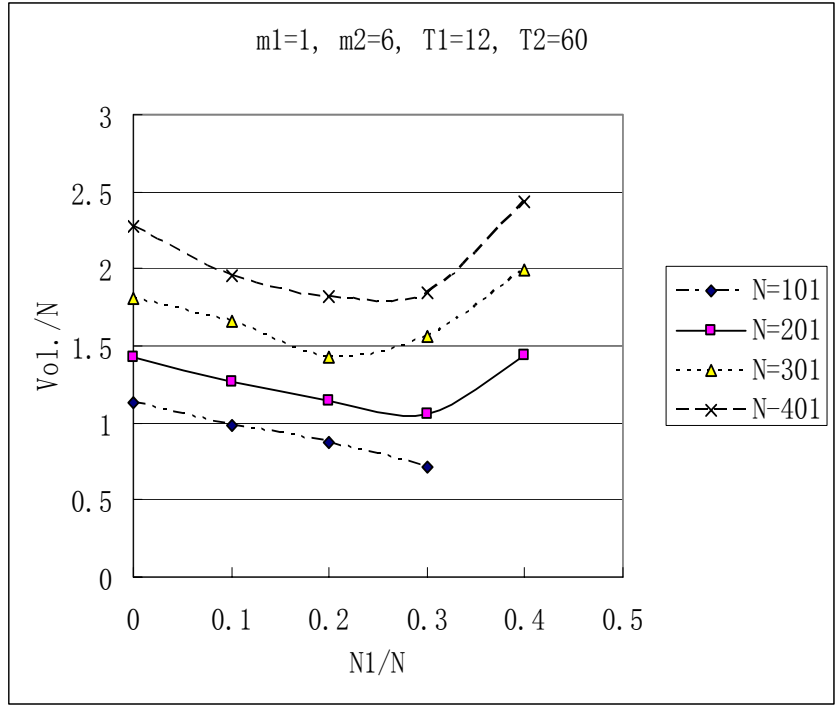

Fig.1a

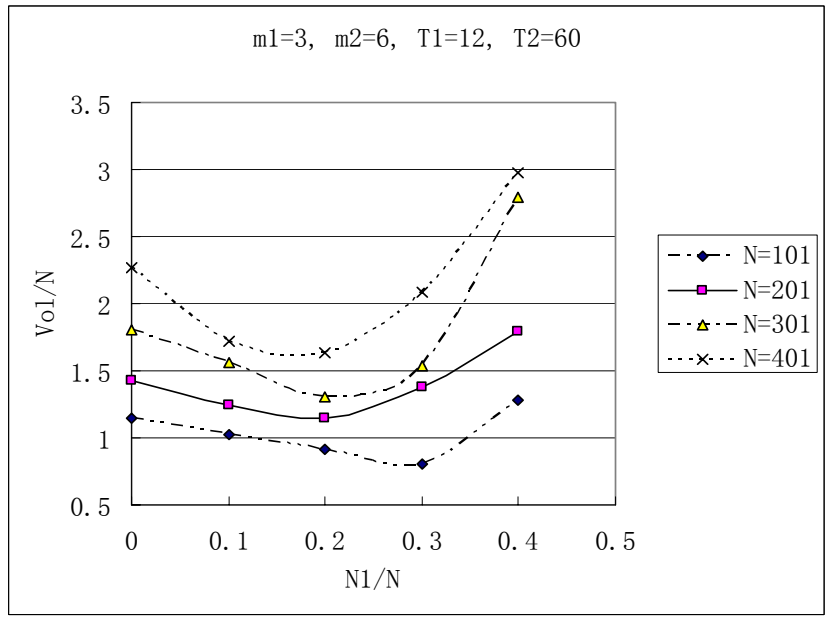

Fig.1b

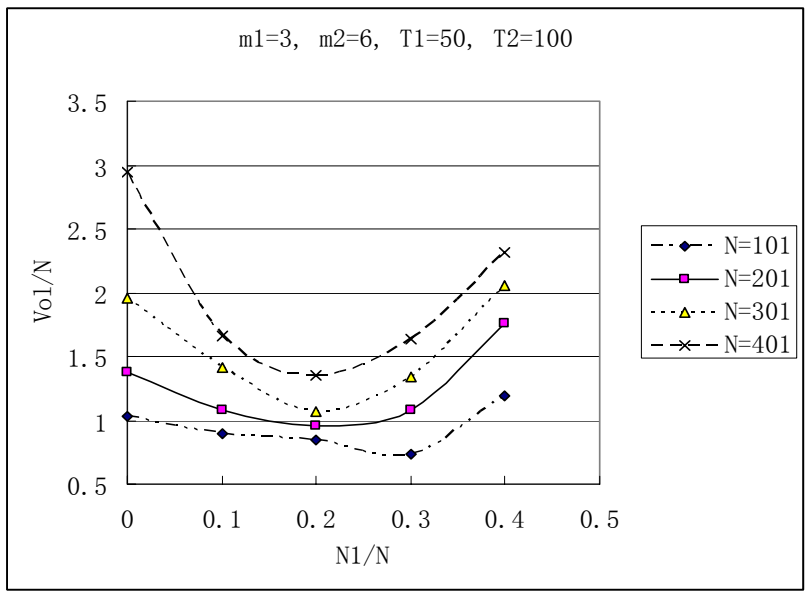

Fig. 1c

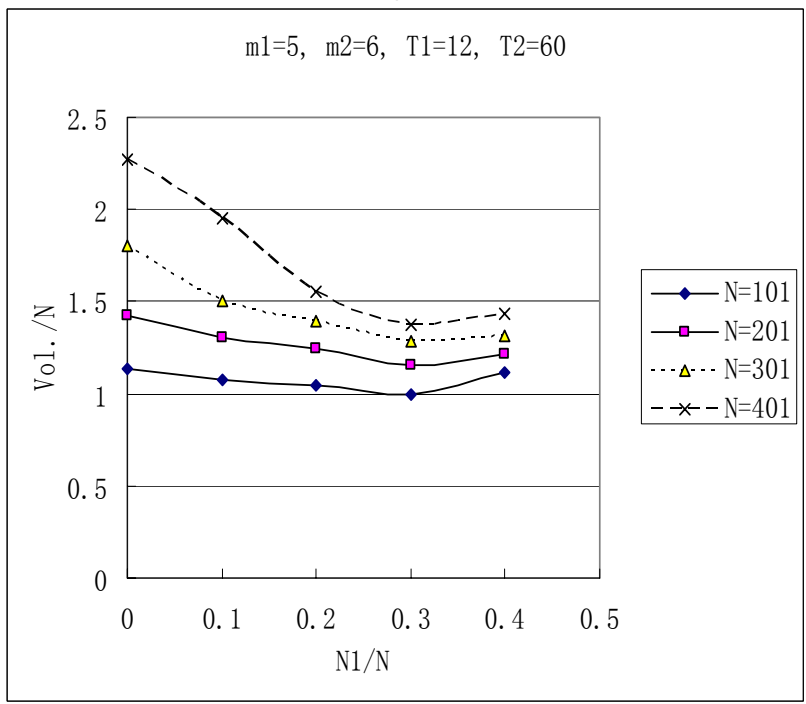

Fig.1d

**Fig. 1**  Vol/N representing the mean local volatilities per agent which is equal to the mean of local volatilities divided by the number of total agents (N) changes with N1/N under the condition of m1<m2, T1<T2 and N from 101 to 401. Fig.1a shows the simulation results with parameters of m1=1, m2=6, T1=12, T2=60; Fig.1b shows the simulation results with parameters of m1=3, m2=6, T1=12, T2=60, and Fig.1c shows the simulation results with parameter m1=3, m2=6, T1=50, T2=100; and Fig.1d shows the simulation results with parameter m1=5, m2=6, T1=12, T2=60

Fig.1 shows the mean local volatilities per agent change when N1/N increases from 0 to 0.4 under the condition of m1< m2=6, T1<T2, N from 101 to 401. We can find that the mean local volatilities per agent have the lowest points at about N1/N=0.2 or 0.3. We can also notice that the mean local volatilities per agent increase while N increase from 101 to 401. Comparing Fig.1b with Fig.1c, we can find that the mean local volatilities per agent are influenced by T1 and T2.  Vol./N in Fig.1d increases less at N1/N=0.4 than those in Fig.1a, Fig. 1b and Fig.1c. Reference [29] found similar results when contrarians are introduced into MG, but the difference is that the lowest point of volatility appears when the fraction of contrarians is 0.4.

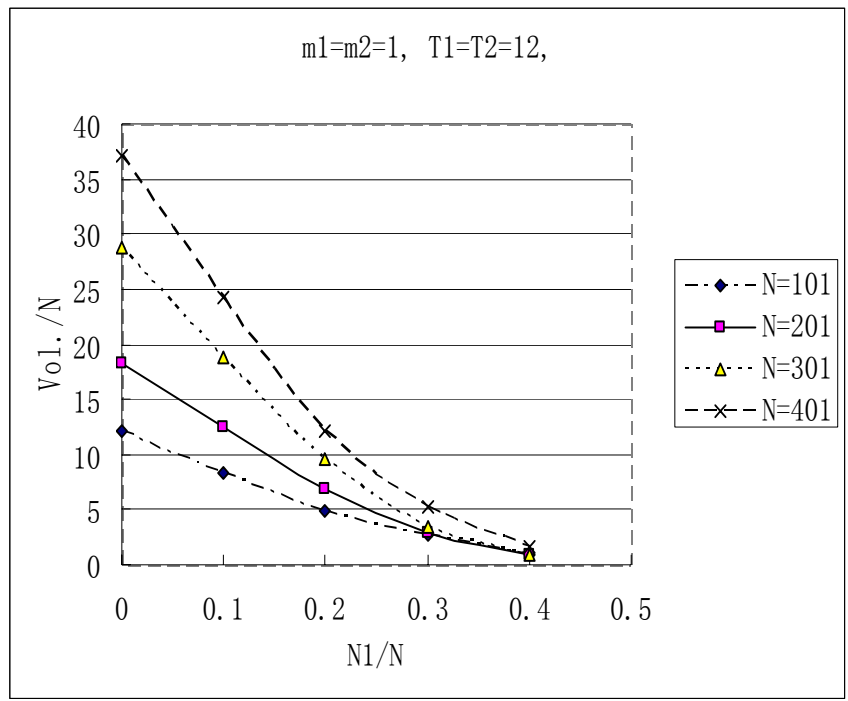

Fig.2a

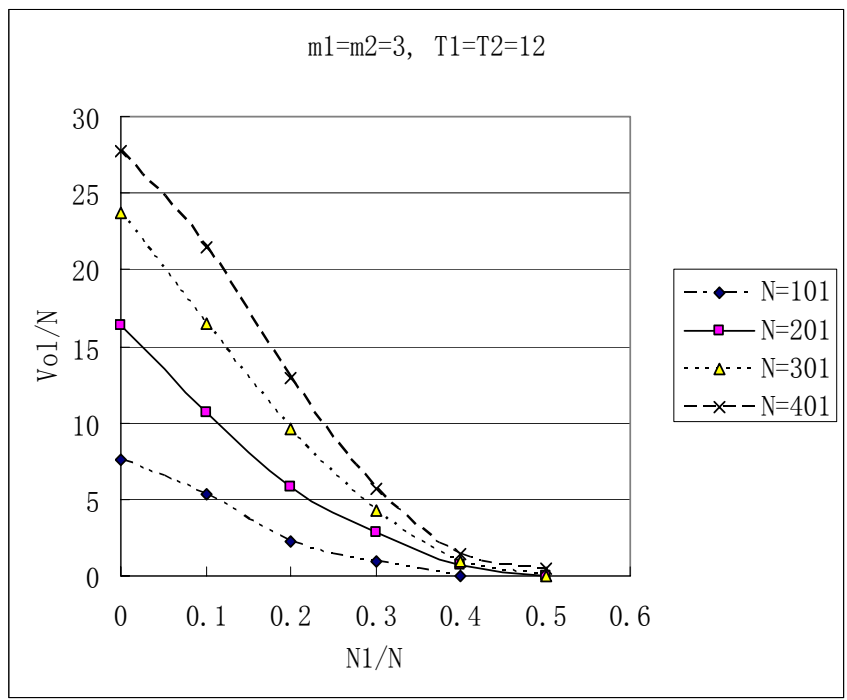

Fig.2b

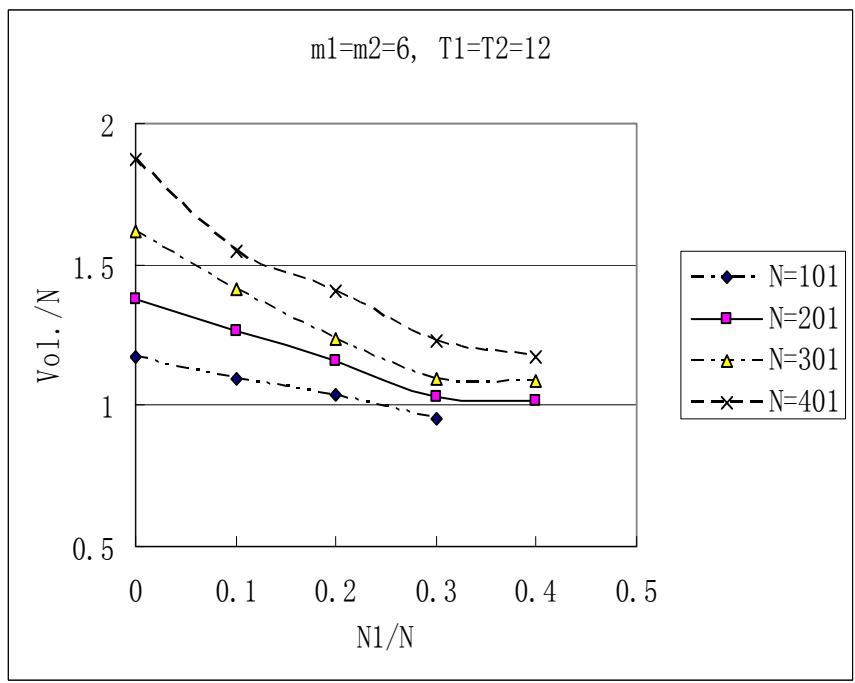

Fig.2c

**Fig. 2** 1 Vol/N representing the mean local volatilities per agent which is equal to the mean of local volatilities divided by the number of total agents (N) changes with N1/N under the condition of m1=m2<=6, T1=T2=12 and N from 101 to 401.

From Fig.2 we can find that the mean local volatilities per agent decrease and approach to 1 while N1/N increase from 0 to 0.4, under the condition of m1= m2<=6, T1=T2=12 and N=101, 201, 301, 401. This means the efficiency of a system can be improved by adding some agents with the same parameters (i.e. m1=m2, T1=T2.) who play the majority game. Reference [20] reported the similar results under the condition of "random information" with N1/N<0.5. And reference [14] observed the similar phenomena in distributed computing systems. Comparing Fig2.a and Fig.2b with Fig.2c, we can find that the mean local volatilities per agent in Fig.2a and Fig.2b are much larger than those in Fig.2c. This means that smaller m1, m2 make the mean local volatilities per agent increase dramatically.

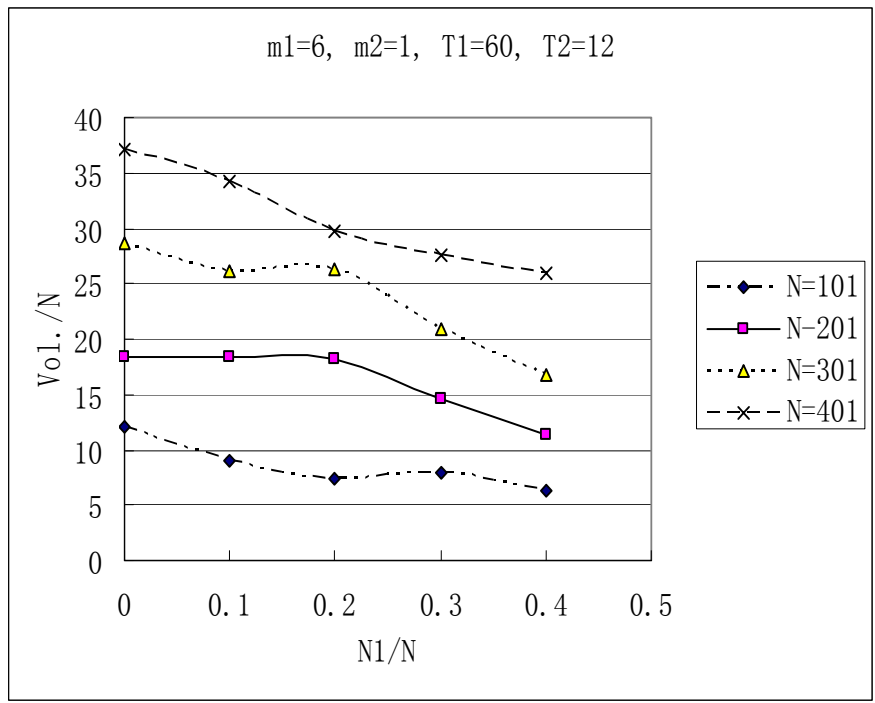

Fig.3a

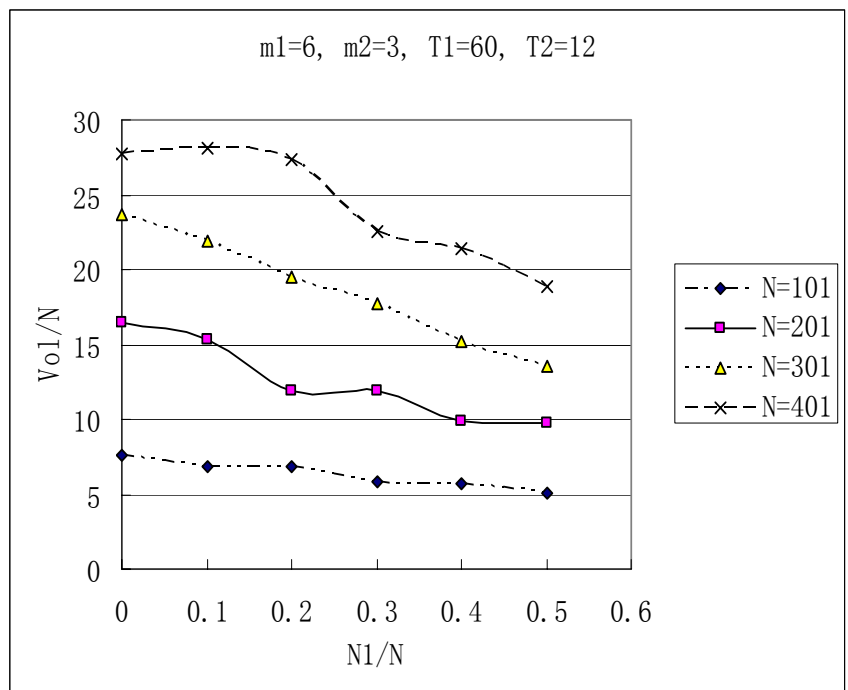

Fig.3b

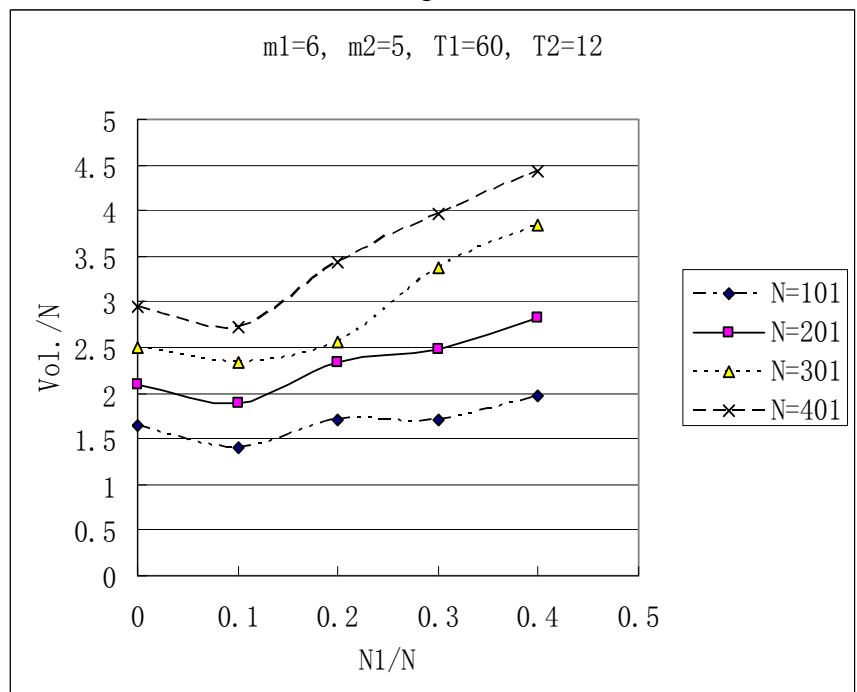

Fig.3c

**Fig. 3** Vol/N representing the mean local volatilities per agent which is equal to the mean of local volatilities divided by the number of total agents (N) changes with N1/N under the condition of m1=6> m2, T1=60, T2=12 and N from 101 to 401.

Fig.3 shows the simulation results with m1=6>m2 and T1>T2, where the mean local volatilities per agent some times decrease, but some times increase while N1/N increases from 0 to 0.4.  For situations of m1=6, m2=1 and m1=6, m2=3, the mean local volatilities per agent decrease when N1/N increases fro 0 to 0.4; for situation of m1=6, m2=5, the mean local volatilities per agent (Vol./N) have the lowest points at N1/N=0.1. However, we can notice that the mean local volatilities per agent increase while N increases from 101 to 401. Comparing Fig.3a, Fig.3b with Fig.2a, Fig.2b, we can find that Vol./N in Fig.3a and Fig.3b decrease much more slowly than those in Fig.2a and Fig.2b when N1/N increases from 0 to 0.4. Comparing Fig.3c with Fig.1d and Fig.2c, we notice that Vol./N in Fig.3c behaves quite differently from those in Fig.1d and Fig.2c.

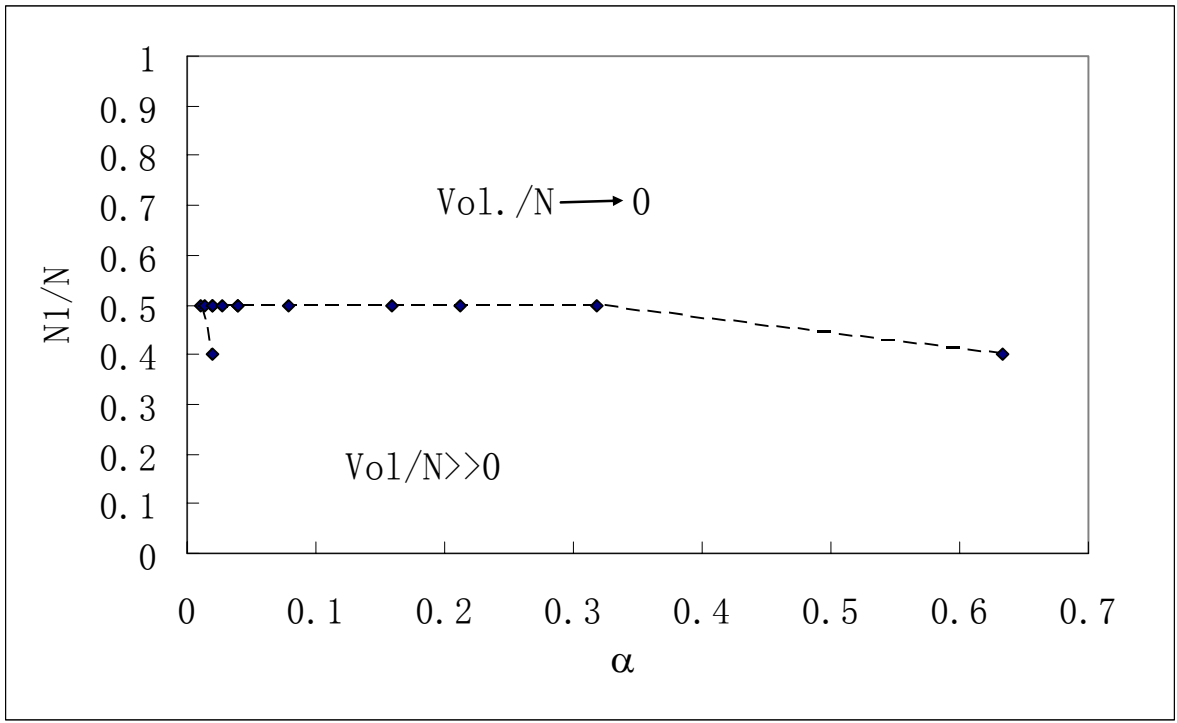

Fig. 4

Fig.4 shows the two-phase phenomenon of local volatilities under simulation condition of m1=m2 or m1<m2, where $\alpha=P/N$, $P=$ power (2, m1).

Through simulations, I find that local volatilities drop to zeros in most cases when N1/N is larger than 0.5 under the simulation conditions of m1=m2 or m1<m2 with some initial strategy distributions. This phenomenon also is influenced by T1. But this phenomenon has not been found under the simulation condition of m1>m2. Fig.4 shows the two-phase phenomenon of local volatilities under simulation condition of m1=m2 or m1<m2, where *α=P/N, P= power (2, m1)*. The reason about local volatilities dropping to zeros is that all agents quickly adapt to using one strategy and stick to it, i.e. all agents are "frozen" in these simulation conditions. Local volatilities drop to zeros can be found frequently in pure majority game, and the smaller the history memory m is, the easier the agents become "frozen". Therefore, we can say that characteristics of minority game dominants in the regime of N1/N <0.5 and characteristics of majority game dominants in the regime of N1/N>=0.5 except two point as shown in Fig.4 where the phases change at N1/N=0.4. However, reference [20] and [29] did not report such phenomenon. A possible explanation for this is that agents receive "random information" for reference [20] and agents reward their strategies according to minority rule for reference [29] so that agents fail to adapt to keep using one strategy.

### 3.2 Analysis and discussion

In order to understand the simulation results, a theoretical analysis can be made as following. First, we consider the situation of m1=m2. Because these two groups have the same memory lengths, i.e. their strategy spaces are the same, and group1 plays majority game and group2 plays minority game, their strategies can be fully anticorrelated if agents in two groups are uniformly distributed in the strategy spaces. Therefore, the dynamic process of mix-game with m1=m2 behaves like the dynamic process of pure minority game with "the effective number of agents" if $x = N1/N < 0.5$. Reference [18] also shows the dynamics behavior of mix-game with m1=m2 is similar to that of pure MG with the same parameters. According to Crowd-Anticrowd theory [2], volatility of MG is proportion to square of the number of agents, i.e. $Vol(x=0) \propto N^2$. So in mix-game with m1=m2, volatility is proportion to square of "the effective number of agents", i.e. $Vol(x) \propto (N^{eff.})^2$. Because the strategies between these two groups are fully anticorrelated, "the effective number of agents" is equal to the number of agents in

group2 minus the number of agents in group1, i.e. $N^{eff.} = N2 - N1$. Since $N2 = (1-x)N$, $N1 = xN$,

we get $N^{eff.} = (1-2x)N$. Therefore, we obtain

$$Vol(x) \propto (N^{eff.})^2 = (1-2x)^2 N^2 \qquad (4)$$

By defining $$y = \frac{Vol(x)}{Vol(x=0)}, \qquad (5)$$

we get $$y = (1-2x)^2 \qquad (6)$$

Fig. 5 is plotted according to formula (6). Comparing Fig.5 with Fig.2, we can find that Fig.2a and Fig.2b qualitatively match Fig.5, but volatility in Fig.2c decreases slowly. This implies that the strategies of group1 are indeed fully anticorrelated with that of group2 in mix-game when memory lengths are m1=m2=1 or 3, and the strategies of group1 is partly anticorrelated with that of group2 when memory length is m1=m2=6. The reason for this is that the strategy space of memory length m=6 is much larger than that of m=1 or 3 so that agents are less uniformly distributed in strategy space of m=6 than that in strategy space of m=1 or 3.

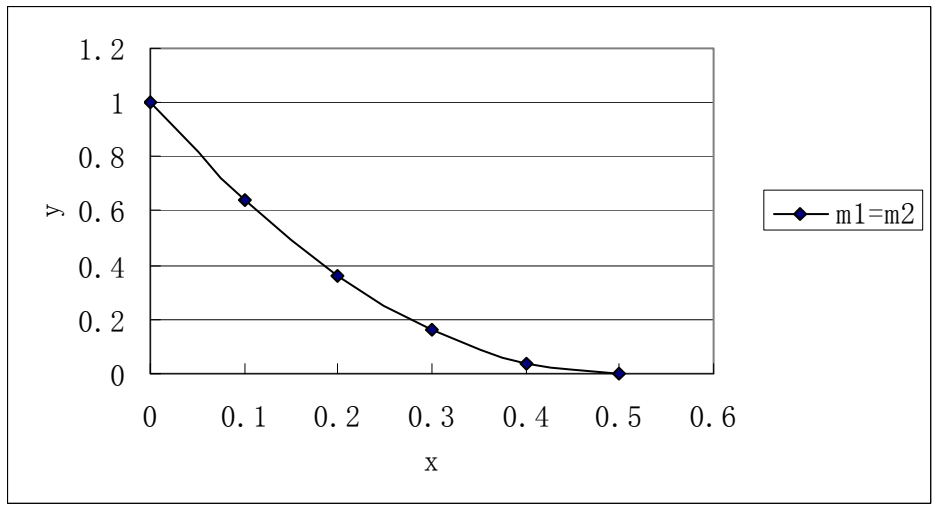

Fig. 5 shows the relation of *y* and *x* according to formula (6), where *x=N1/N*.

Second, we look at the situation of m1=6>m2. Because the strategy space of group1 is much larger than that of group2, the distribution of agents in strategy space 1 is not uniform so that the strategies held by

group1 have little chance to anticorrelate with that held by group2. Therefore, we can assume that these two groups behave independently. The volatility can be calculated as

$$Vol(x) = Vol_{m1=6}(x) + Vol_{m2}(x). \qquad (7)$$

So we get 
$$y = \frac{Vol(x)}{Vol(x=0)} = \frac{Vol_{m1=6}(x) + Vol_{m2}(x)}{Vol_{m2}(x=0)}. \qquad (8)$$

Through simulation, I obtain $Vol_{m1=6}(x)$, $Vol_{m2}(x=0)$ and $Vol_{m2}(x)$. Fig.6 is plotted according to formula (8) with N=401, m1=6, m2=1, 3, 5, T1=T2=12. Comparing Fig.6 with Fig.3, we can find volatilities in Fig.6 qualitatively match with that in Fig.3. This shows that the argument about that these two groups behave independently is reasonable. Reference [31] also supports this point.

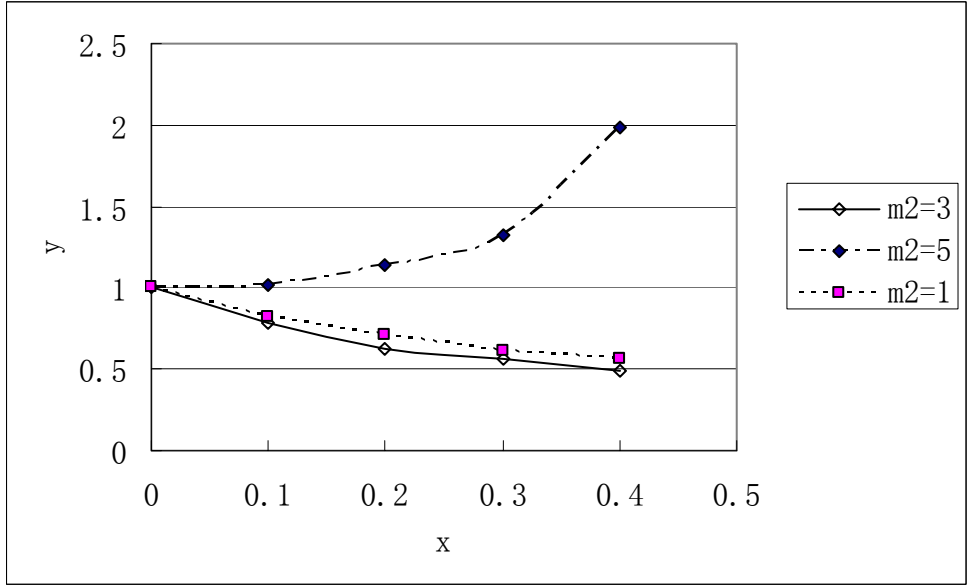

Fig.6 shows the relation of *y* and *x* according to formula (8), where *x=N1/N*.

Finally, we discuss the situation of m1<m2=6. Because the strategy space of group1 is much lsmaller than that of group2, the distribution of agents in strategy space 1 is relatively uniform so that the strategies held by group1 have chance to partly anticorrelate with that held by group2. I refer to this as "anticorrelated effect". On the other hand, through simulations I find that the volatilities of pure majority game are several times larger than that of pure minority game with the same parameters. I refer to this as "majority game effect". Therefore, the dynamic processes of mix-game with m1<m2=6 are influenced by the interplay between the anticorrelated effect and the majority game effect. This is why the volatilities have the minimums at $x_{min}$. If $x < x_{min}$, the anticorrelated effect dominants in mix-game. If $x > x_{min}$, the majority game effect dominants in mix-game. Mix-game in this situation is more interesting than it in the previous two situation because mix-game with m1<m2=6 has potential applications in modeling financial markets. I will discuss this point in the following section.

## 4. Application of mix-game model

**4.1 The effect of mix-game on local volatilities**

In order to see if mix-game is a potential model for financial markets, one needs to know the effect of mix-game on time series of prices. Therefore, I examine the time series of prices of MG and mix-game with the same parameters as that in the above section. I find that agents' playing majority game nearly do not change the stylized features of price time series of MG except m1=1 or 2 <m2=6 if N1/N<0.5. One naïve explanation for this is that minority game dominates in mix-game when N1/N<0.5. Fig.7 shows the examples of the time series of prices which are done under condition of N=201, N1=72, T1=T2=12, m1=1, 2, 3<m2=6 in mix-game.

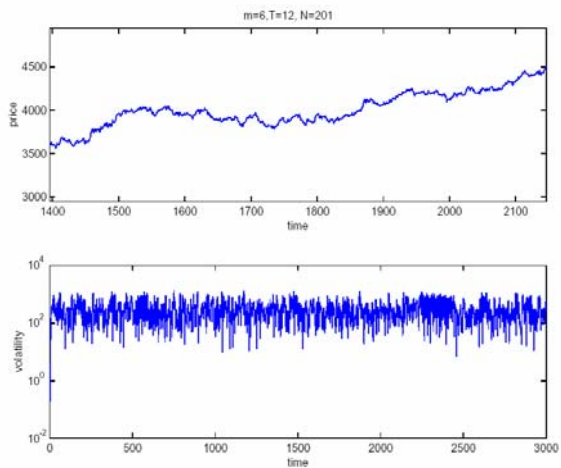

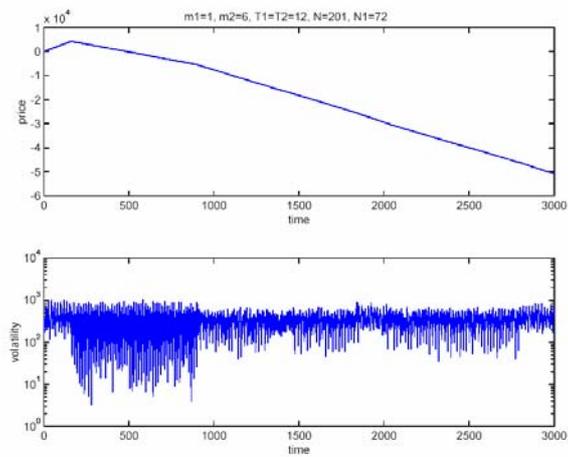

Fig. 7a                                  Fig. 7b

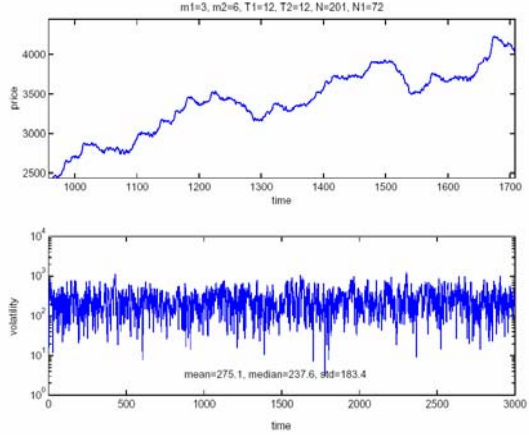 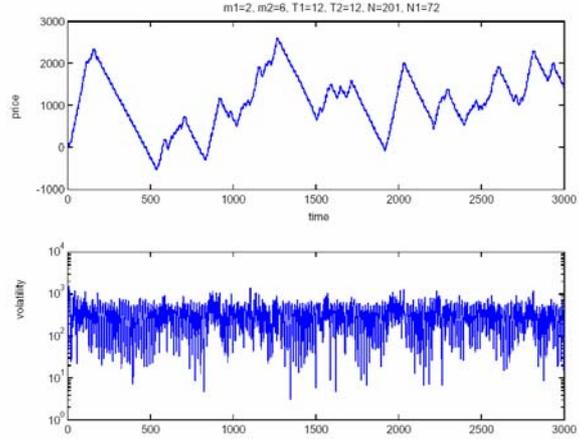

Fig. 7c                                                                 Fig. 7d

**Fig. 7** time series of prices and local volatilities of MG with m=6, T=12 and N=201, and mix-game with T1=T2=12 , N=201, N1=72, m1=1, 2, 3<m2=6;  the up half  figures are price time series and the down half figures are time series of local volatilities and the x-axes represent time steps.

Table 1 the average winnings of both groups in mix-game and MG where N=201, N1=72, T1=T2=T=12

|  | Mix-game | | MG |
|---|---|---|---|
|  | m1=1, m2=6 | m1=2, m2=6 | m=6 |
| Group1 | 0.693 | 0.612 | 0.466 |
| Group2 | 0.511 | 0.493 |  |

Comparing with the price time series of MG with m=6, T=12 and N=201, one can find that the features of price time series in mix-game nearly don't change if some agents with 6>=m1>=3 play the majority game and other agents with m2=6 play minority game. Fig7c is an example of this phenomenon. However, the features of price time series change dramatically if some agents with m1=1 or 2 play the majority game under condition of m2=6, as shown in Fig.7b and Fig7d.  The price time series show strong trends, which do not appear in real price time series of financial markets. J. V. Andersen and D. Sornette already noticed the similar phenomenon for the $-game in reference [4]. In reference [18], I find that the agents' average winnings in mix-game under the condition of m1=1 or 2, m2=6, T1=T2=12, N=201, N1=72 are much bigger than that in MG under the condition of m=6, N=201, T=12, as shown in Table 1. This result implies that agents in mix-game who are not only in majority group but also in minority group can take the advantage of the trends of price time series to improve their performances.

**4.2 The rules and the example for simulating financial markets by mix-game**

From the former sections, one can find out that mix-game with some parameters reproduces the stylized features of financial time series, but some fail to do so. Therefore, one needs to choose parameters of m1, m2, T1, T2, N and N1 by using mix-game to model financial markets. The following aspects need to be considered:

- First make sure the time series of mix-game can reproduce the stylized facts of price time series of financial markets by choosing proper parameters, and m1 and m2 must be larger than 2;
- Second pay attention to the fluctuation of local volatilities, and ensure that the median of local volatilities of mix-game is similar to that of the target time series;
- Third make sure the log-log plot of absolute returns look similar.

Since the median value of local volatilities of Shanghai Index daily data from 1992/01/02 to 2004/03/19 is 222, two combinations of parameters of mix-game have the similar median values of local volatilities according to Fig.4; one combination is m1=3, m2=6, T1=12, T2=60, N=201, N1=40, the other is m1=4, m2=6, T1=T2=12, N=201, N1=72.

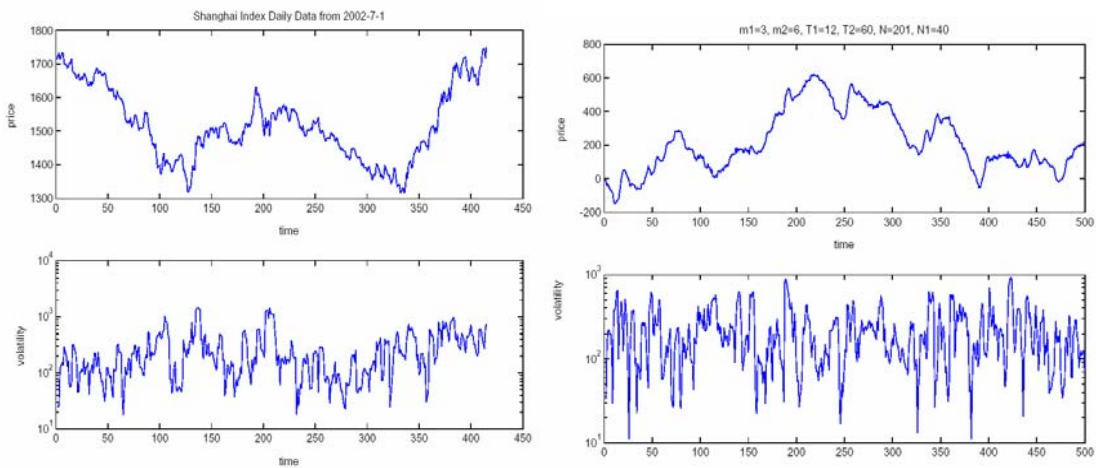

**Fig. 8** time series and local volatilities of shanghai index daily data and mix-game with parameters of m1=3, m2=6, T1=12, T2=60, N=201, N1=40

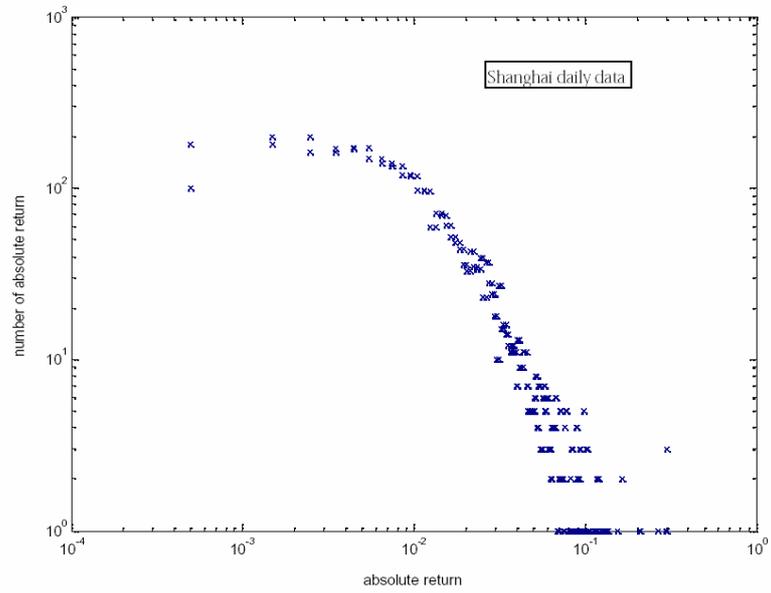

**Fig. 9.** log-log plot of shanghai index daily absolute returns which is non-Gaussian [26].

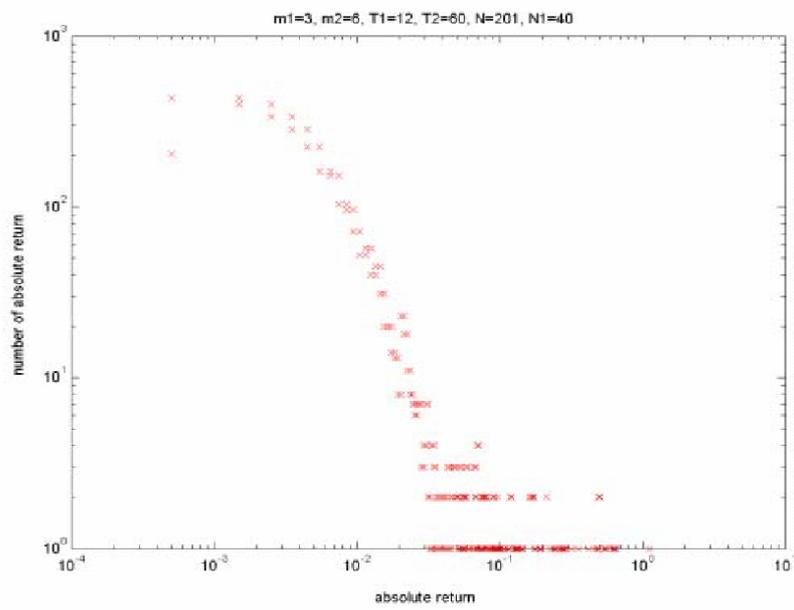

**Fig.10** log-log plot of mix-game absolute returns with parameters of m1=3, m2=6, T1=12, T=60, N=201 and N1=40

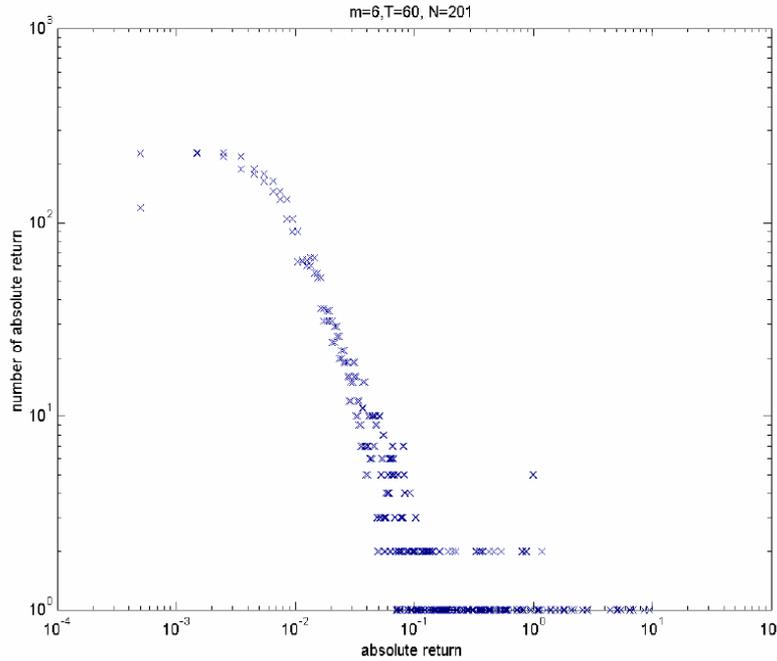

**Fig.11** log-log plot of MG absolute returns with parameters of m=6, T=60 and N=201

Fig.8, Fig.9 and Fig.10 show that mix-game with m1=3, T1=12, m2=6, T2=60, N1=40, N=201 can reproduce the similar stylized facts of Shanghai stock market. So does the combination of parameters of m1=4, m2=6, T1=T2=12, N=201, N1=72. As shown in Fig.11, MG with parameters of m=6, T=60 and N=201 has a much fatter tail in log-log plot of its absolute returns than that of Shanghai index daily-data. So for simulating Shanghai Index, mix-game behaves better than MG.

## 5. Conclusion:

In mix-game, different configurations of agents' history memory lengths largely influence the local volatilities: means of local volatilities decrease monotonously with increase of N1/N from 0 to 0.4 if m1=m2=1, 3 or 6; means of local volatilities decrease first and then increase with increase of N1/N from 0 to 0.4 if m1<m2=6; means of local volatilities sometimes increase and sometimes decrease if m1=6>m2 with increase of N1/N from 0 to 0.4. It is found that local volatilities drop to zeros in most cases when N1/N is larger than 0.5 under the simulation conditions of m1=m2 or m1<m2 with some initial strategy distributions.

The underlying mechanisms for the above findings are as following: if these two groups have the same history memories, their strategies are anticorrelated; if history memory of group1 is 6 and larger than that of groups 2, their strategies are uncorrelated; if history memory of group2 is 6 and larger than that of groups 1, their strategies are partly anticorrelated.

Besides, history memory lengths, m for MG, m1 and m2 for mix-game influence the dynamics of price time series. Finite time horizons T for MG, T1 and T2 for mix-game also have influence on the dynamics of price time series, as studied by Hart and Johnson [27] and D. Chalet [28]. Therefore, further study will focus on the combined effects caused by changing time horizons and history memory lengths

Mix-game can be a potentially good model for financial markets with specific parameters. For Shanghai Index, there are two suitable configurations of parameters: m1=3, T1=12, m2=6, T2=12, N1=40, N=201 and m1=4, m2=6, T1=T2=12, N=201, N1=72, respectively. For any other financial market, parameters need to be adjusted.

**Acknowledgements**

This research is supported by Chinese Overseas Study Fund. Thanks Professor Neil F. Johnson for helpful discussion. Thanks David Smith for providing the original program code of MG. I appreciate the referees' comments.